\documentclass[aps,pre]{revtex4-2}
\usepackage{amssymb}
\usepackage{amsmath,bm}
\usepackage{graphicx,color}
\usepackage{epsfig}
\usepackage{multirow}
\usepackage[cal=pxtx]{mathalfa}
\usepackage{accents}
\DeclareMathAccent{\wtilde}{\mathord}{largesymbols}{"65}
\usepackage{stackengine}
\usepackage{ulem}
\stackMath
\newcommand\tenq[2][1]{%
	\def\useanchorwidth{T}%
	\ifnum#1>1%
	\stackunder[0pt]{\tenq[\numexpr#1-1\relax]{#2}}{\scriptscriptstyle\sim}%
	\else%
	\stackunder[1pt]{#2}{\scriptscriptstyle\sim}%
	\fi%
}

\begin{document}

\title{Torque Hyperuniformity in Frictional Granular Matter - Theory and Experiments}

\author{Jin Shang$^{*1,2}$ and Jie Zhang$^{2,3}$ and Itamar Procaccia$^{1,4,5}$}
\affiliation{$^1$Hangzhou International Innovation Institute, Beihang University, Hangzhou 311115, China \\$^2$School of Physics and Astronomy, Shanghai Jiao Tong University, Shanghai 200240, China\\$^3$Institute of Natural Sciences, Shanghai Jiao Tong University, Shanghai 200240, China\\$^4$ Sino-Europe Complex Science Center, School of Mathematics, North University of China, Shanxi, Taiyuan 030051, China.\\$^5$Department of Chemical Physics, the Weizmann Institute of Science, Rehovot 7610001, Israel.}

\begin{abstract}
A question of some fundamental importance is whether a given assembly of frictional granules (say sand or powder) will exhibit stress autocorrelations with long-range anisotropic decay as determined by the elastic Green's function. In Hamiltonian systems with central forces, mechanical balance and material isotropy demand the stress auto-correlation matrix to be fully determined by the pressure auto-correlation only. If the local pressure fluctuations are normal, it follows that stress autocorrelations decay at long distance like the elastic Green's function. With friction, Hamiltonian symmetry is lost, and one may expect more constraints. Indeed, it was shown recently that for frictional amorphous solids mechanical balance and material isotropy demand the stress auto-correlation matrix to be fully determined by {\bf two}
spatially isotropic functions: the pressure
{\bf and} torque auto-correlations. Elastic-like decay of the stress autocorrelations follows from normal fluctuations of the pressure {\bf and} from the torque fluctuations being hyperuniform. The theoretical discovery of these conditions required experimental confirmation, to test whether these conditions are generically obeyed in actual frictional amorphous solids. Recently the confirmation was announced for 2-dimensional amorphous assemblies of frictional disks under isotropic load, in which torque is caused by tangential forces only. In this paper we review that case and report confirmation of the theoretical predictions in 2-dimensional systems of disks under shear and in isotropically loaded frictional ellipses, where contributions to torque come also from normal forces. The paper ends with physical explanations of the hyperuniformity of the torque fluctuations and predictions for how the results are expected to extend to d-dimensions. 

\end{abstract}

\maketitle
\section{Introduction}
Understanding the mechanical properties of amorphous solids requires developing many new concepts, approaches and methods. Amorphous solids differ from classical elastic media in being prone to plastic deformations \cite{10KLP,11HKLP}, they also differ from crystalline solids in having excess low-frequency modes (Boson peak) \cite{03GPS}, more than expected from the Debye theory of long-wavelength modes. The discovery of topological transitions in amorphous solids leading to anomalous responses came as a surprise \cite{21LMMPRS,24JPS,25FJPP}. Full understanding of instabilities like shear banding and fracture is still being worked out \cite{14Fie}.

A question of immediate interest regarding amorphous solids in general and granular solids in particular, is whether the long-distance decay of stress autocorrelation functions is expected to follow predictions of elasticity theory in the form of the elastic Green's function. For amorphous solids  whose inter-particle forces derive from a Hamiltonian this question was settled in the last decade  \cite{09HC,14Lem,15Lem,17Lem,18Lem}.  It was demonstrated that in Hamiltonian problems the ``normal" decay of
long range correlations follows from the conjunction of three properties. These are
(i) Mechanical balance, (ii) Material isotropy and (iii) The normality of local pressure fluctuations \cite{17Lem,18Lem}.
The derivation of these results depends crucially on the symmetry of local stress which
inevitably breaks down in the presence of frictional forces which introduce local torques. Therefore until recently the question remained 
fully open about the nature of stress correlations in frictional granular packings, an important, diverse and widespread class of materials including sand, soils, powders etc.

To reiterate, in Hamiltonian systems with central forces, mechanical balance and material isotropy demand the stress auto-correlation matrix to be fully determined by the pressure auto-correlation only. The situation in frictional granular packing is richer. In a recent theoretical paper \cite{25SZP} it was shown that in  frictional granular packings,  the stress auto-correlation matrix is determined not by one but by two spatially isotropic
functions, the pressure and torque autocorrelations. Moreover, elastic-like decay of the stress autocorrelation functions requires again normal pressure fluctuations, but the torque fluctuations must be hyper-uniform, i.e. the torque auto-correlation vanishes in the
zero wave-number limit. As a consequence the torque contribution to the stress auto-correlation is
sub-dominant at large wave-length. Consequently, the large distance decay of the stress-autocorrelation
is again determined by the scaling of local pressure fluctuations on domains of increasing sizes. When these
fluctuations are normal the presence of elastic-like long-ranged anisotropic contributions follows.

The theoretical statement that torque fluctuations must be hyperuniform does not mean that actual frictional matter will always bow to this requirement. To judge whether this condition is universally obeyed we must turn to experiments. For such experiments to be successful one needs to measure the forces on each component, both normal and tangential, to be able to evaluate the pressure and the torque. This limits the choices, and we examine the issue using granules made from plastic material that rotate polarized light when strained, allowing to measure all the forces. Recently we reported that in the case of plastic disks, in which all the contribution to the tangential forces
comes from frictional forces, the hyperuniformity of the torque fluctuations was confirmed \cite{25SZP}. Here we review these results and add two more systems, one of disks under shear and the second  made from plastic ellipses, in which tangential forces form due to friction and a contribution from the normal force. We conclude below that also in these
systems the hyperuniformity of the torque fluctuations is confirmed.

The structure of this paper is as follows: in Sec.~\ref{hyper} we clarify the notions of normal and hyperuniform fluctuations for the reader who is not familiar with these terms. Sec.~\ref{expdisc} describes the experimental apparatus and procedure for the case of disks, followed by the experimental results and interpretations. In Sec.~\ref{ell} we turn to the experiments using
plastic ellipses, and provide experimental details and results for this case. In Sec.~\ref{summary} we summarize the results, provide intuitive arguments to understand the experimental results, and generalize to predict how torques fluctuations are expected to scale in d-dimensions.  

\section{Normal and Hyperuniform Fields}
\label{hyper}

 The term hyperuniformity was coined by Torquato and Stillinger in Ref.~\cite{03TS}, and it was found useful in many scientific contexts, from Solid State Physics \cite{03TS} and  Cosmology \cite{02GJS} all the way to Biology \cite{14JLHMCT}. Hyperuniformity refers to the behavior of the fluctuations of a density, be it a number density, a vector or tensor density, at large scales. Denoting the density as $q$, and the integral of $q$ over a d-dimensional ball of radius
 $R$ as $Q$, one examines how the variance of $Q$ grows when computed in a sphere of increasing radius $R$. ``Normal" fluctuations exhibit variance that grows like $R^d$ in $d$-dimensions. When the variance grows slower than $R^d$ for large $R$, the density is referred to as hyperuniform. Precisely, a density Q is hyperuniform when  
\begin{equation}
	\lim_{R\to \infty} \frac{	\langle (Q-\langle Q\rangle)^2\rangle }{R^d} =0 \ . 
\end{equation}
Here $\langle \cdots\rangle$ denotes an average over realizations.  A consequence of this property is the vanishing of the structure factor at the origin:
\begin{equation}
	\lim_ {k \to  0} S ( \mathbf{k} ) = 0 \quad \text {for wave vectors} ~\mathbf{k} \in R^d  .
\end{equation}
We will see below that in the three 2-dimensional experimental systems discussed in this paper, the variance of the torque fluctuations increase like $R$. An explanation of this scaling law is provided in Sec.~\ref{summary}

\section{Experiments with disks} 
 \label{expdisc}
 
 We studied assemblies of frictional disks under two types of load, isotropic compression and pure shear. Results of the isotropic compression were announced recently \cite{25SZP} and are included here for completeness. The results for pure shear are new to this paper. We note that pure shear tends to destroy the system's isotropy, one of the conditions for elastic like decay of the stress autocorrelation functions. Nevertheless we discover below that the torque fluctuations remain hyperuniform. 
\subsection{Experimental Setup for isotropic compression}
\label{esic}
\begin{figure}
	\centering
	\includegraphics[width= 8.6 cm]{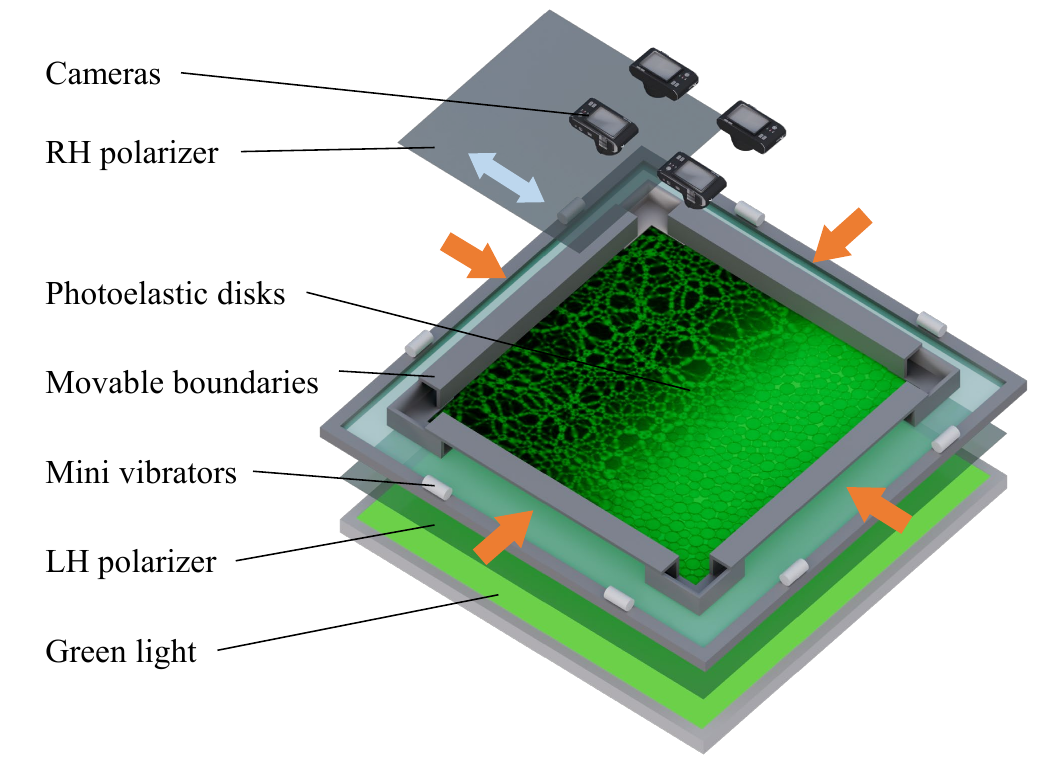}
	\caption{Schematic of the experimental setup. Inside we show a typical image in which half the cell exhibits disks and the other half shows force chains.}
	\label{apparat}
\end{figure}

Circular disks under isotropic load were examined using 2710 small and 1350 large photoelastic disks (in fact, flat cylinders), made of Vishay PSM-4, with diameters of 10 mm and 14 mm, respectively.  The experimental apparatus, cf. Fig.~\ref{apparat}, is composed of a biaxially movable square frame mounted on a horizontal glass substrate.  The sidewalls of the square frame are capable of inward movement at a speed of 0.1 mm/s, thereby imposing quasi-static isotropic compression on the inner granular packing. 
Anisotropic loading (simple shear) is considered in Subsect.~\ref{pure}. To minimize the influence of friction between the disks and the underlying glass substrate, the glass surface is coated with a thin layer of talcum powder. Additionally, mini vibration motors, affixed to the aluminum frame supporting the glass plate, are activated during boundary motion. 

To generate a jammed packing at a specific pressure, an initial stress-free, random, and homogeneous configuration is prepared at a packing fraction $\phi = 0.834$ slightly below the jamming point $\phi_J\approx 0.84$ of the corresponding frictionless configuration\cite{OHern2002prl}. Subsequently, quasi-static isotropic compression is applied step by step until the target packing fraction $\phi = 0.862$ is reached, corresponding to an average pressure of $p = 31.1 \ \mathrm{N\,m^{-1}}$. This preparation protocol yields force networks that are isotropic and approximately homogeneous. In the description of the experimental results below we choose the radius of the small disk as the unit of length. To mitigate potential boundary effects, statistical analyses exclude about 10 layers of disks near the boundaries.

A green LED light source is positioned beneath the apparatus, overlaid with a left-handed circular polarizer. Particle configurations and stress distributions are captured via a $2\times2$ array of high-resolution (10 pixels/mm) digital cameras, located about 1.5 m above the apparatus.  A movable right-handed circular polarizer is installed in front of the cameras. When the polarizer is removed, normal images are captured to determine the spatial positions of the disks. Upon insertion of the polarizer, photoelastic stress images are obtained, and a force-inversion algorithm \cite{05MB,Wang2020nc,Wang2021prr} is used to reconstruct the vector contact forces between the disks.
The torque and pressure are defined at each contact, without any coarse graining.

\begin{figure}
	\centering
	\includegraphics[width= 8.6cm]{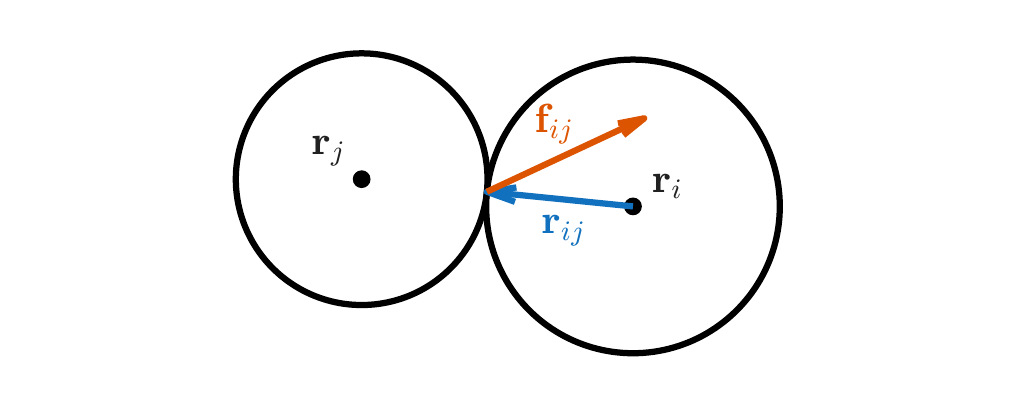}
	\caption{Definitions of $\mathbf{r}_i$, $\mathbf{r}_{ij}$ and $\mathbf{f}_{ij}$.}
	\label{definition}
\end{figure}
\subsection{Data Analysis}
\label{anal}
 The torque and pressure fields can be expressed as
\begin{equation}
	\tau(\mathbf{r}) = -\sum_{i,j}(\mathbf{r}_{ij} \times \mathbf{f}_{ij}) \cdot \hat{\mathbf{z}} \delta [\mathbf{r} - (\mathbf{r}_i + \mathbf{r}_{ij})],
  \label{eq:tau}
\end{equation}
\begin{equation}
	p(\mathbf{r}) = -\sum_{i,j} \mathbf{r}_{ij} \cdot \mathbf{f}_{ij} \delta [\mathbf{r} - (\mathbf{r}_i + \mathbf{r}_{ij})],
  \label{eq:pressure}
\end{equation}
where the definitions of the variables are shown in the Fig.~\ref{definition}. The unit vector  $\hat{\mathbf{z}}$ is orthogonal to the plane of the disks.

\begin{figure*}
	\centering
	\includegraphics[width= 14 cm]{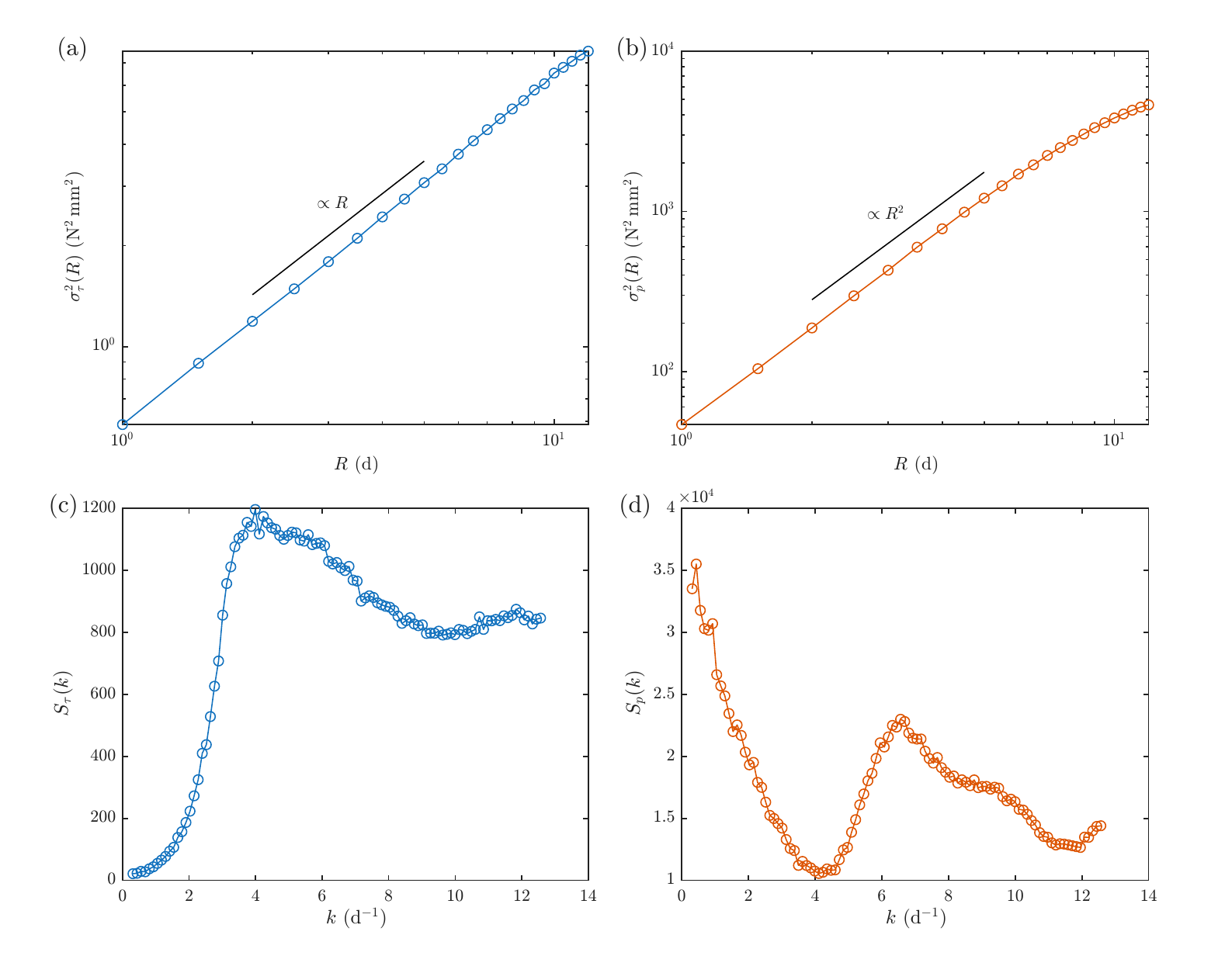}
	\caption{(a) The $R$ dependence of the variance of of torque fluctuations. The linearity in R is a direct evidence of hyperuniformity. Below we explain that the result is exact, i.e. the growth of variance of the torque is like $R^\alpha$ with $\alpha=1$. (b) The $R$ dependence of the variance of of pressure fluctuations. (c) $S(k)$ of torque. (d) $S(k)$ of pressure.}
	\label{FluctSK}
\end{figure*}

 To assess the nature of the fluctuations of our two fundamental fields,  we integrate the torque (or pressure) within a circular window of a random center $\mathbf{r}$ and a radius $R$. The sampling is repeated 4000 times for each window radius and then the variance of these integrals are calculated. For torque, it is
\begin{equation}
	\sigma_{\tau}^2(R) = \mathrm{Var} \left ( \sum_{|\mathbf{r}_{i} + \mathbf{r}_{ij} - \mathbf{r} |<R} (\mathbf{r}_{ij} \times \mathbf{f}_{ij}) \cdot \hat{\mathbf{z}} \right ),
  \label{eq:var}
\end{equation}
and similarly for the pressure. 

\subsection{Results}

As shown in the Fig.~\ref{FluctSK} (a) and (b), the variance of the torque fluctuation increases like $R$ and for pressure like $R^2$. The deviations at large scales are due to the finite-size effect. 

Finally, we examine the Fourier spectra of our field fluctuations.  
The spectrum of the torque fluctuations is defined (with a similar definition for the pressure) as
\begin{equation}
	S_{\tau}(\mathbf{k}) = \left | \sum_{i,j}\delta \tau_{ij} \mathrm{e}^{-\mathrm{i} \mathbf{k} \cdot (\mathbf{r}_i + \mathbf{r}_{ij})} \right |^2
  \label{eq:sk}
\end{equation}
with $\tau_{ij} = (\mathbf{r}_{ij} \times \mathbf{f}_{ij}) \cdot \hat{\mathbf{z}}$ and $\delta \tau_{ij}  = \tau_{ij} - \langle \tau_{ij} \rangle$.
After an angular averaging each spectrum on $\mathbf{k}$, and then averaging over 30 independent samples, the results are shown in Fig.~\ref{FluctSK} (c) and (d). The vanishing of the spectrum of the torque fluctuations at $k\to 0$ is a direct evidence of hyperuniformity.

\subsection{Experimental setup for pure shear}
\label{pure}

The experimental apparatus and the disk particles used are identical to those described in Sec.~\ref{esic}. Isotropic initial states are first prepared at various packing fractions $\phi$ using the compression protocol from Sec.~\ref{esic}. We then apply area-conserving pure shear by moving the x-boundaries inward and the y-boundaries outward at a slow rate ($0.2\, \mathrm{mm\, s^{-1}}$). The shear is applied in discrete steps, in which the boundaries in the x-direction are each displaced by $0.4\, \mathrm{mm}$, with data acquisition after each step. 

 Because shear dilatation causes the pressure to increase with strain, a single initial state cannot cover a wide range of strains while remaining within our optimal pressure window for force measurement (about $15-30\, \mathrm{N\,m^{-1}}$). To address this, and to allow for comparisons of different strains at the same pressure, we perform experiments starting from six different initial packing fractions (and thus initial pressures), ranging from 0.843 to 0.866. For each case, 20 independent experimental runs are performed.
\subsection{Data Analysis and Results}
\begin{figure}
	\centering
	\includegraphics[width = 14 cm]{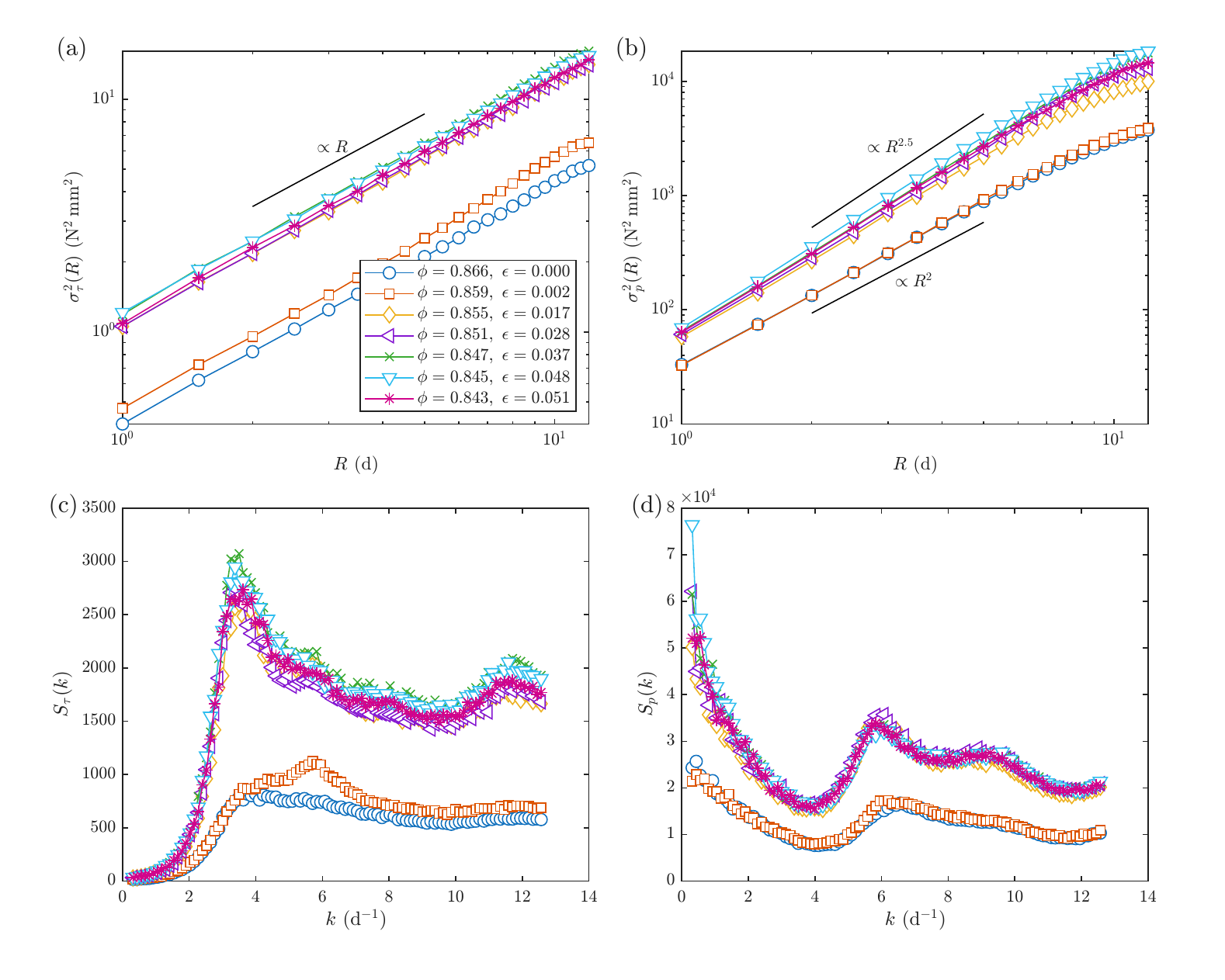}
	\caption{$\sigma^2(R)$ and $S(k)$ for systems at similar pressure $p\approx 25\ \mathrm{N\,mm^{-1}}$ with different packing fractions $\phi$ and strains $\epsilon$.}
	\label{fig:f2}
\end{figure}

The data analysis for pure shear follows verbatim the definitions and calculations presented in Subsect~\ref{anal}. We should note however that  due to friction, pure shear loading with constant area leads to a significant increase in pressure. To compare systems at the same pressure, we varied the initial pressure $p$ of the un-sheared state by changing the system's packing fraction $\phi$. This allowed us to obtain systems with similar pressures at different strains $\epsilon$. Here, the strain is defined as the displacement of the system's compression axis divided by its original length, $\epsilon = \Delta x/x_0$.

Fig. \ref{fig:f2} shows the results for different packing fractions and strains at a pressure of approximately $p \approx 25\, \mathrm{N\,mm^{-1}}$, with an isotropic system ($\epsilon = 0$) included for comparison. We can see that beyond a certain strain, possibly $\epsilon \approx 0.017$, the behavior becomes nearly invariant.

For the torque $\tau$, the fluctuations consistently exhibit hyperuniformity. Although the fluctuation values increase after shearing, the shape of the $S(k)$ curve also changes significantly.

For the pressure $p$, the fluctuations shift from normal uniform fluctuations to "hyperfluctuations" after shearing, with $\sigma_p^2\sim R^{2.5}$. This is due to the loss of isotropy, leading to  heterogeneous pressure distribution. Correspondingly, we also observe a diverging trend in $S_p(k)$ at small $k$. These effects were observed and explained in Ref.~\cite{21LMPRWZ}, but here we focus on the 
torque fluctuations which remain robustly hyperuniform. 

\begin{figure}
	\centering
	\includegraphics[width = 14 cm]{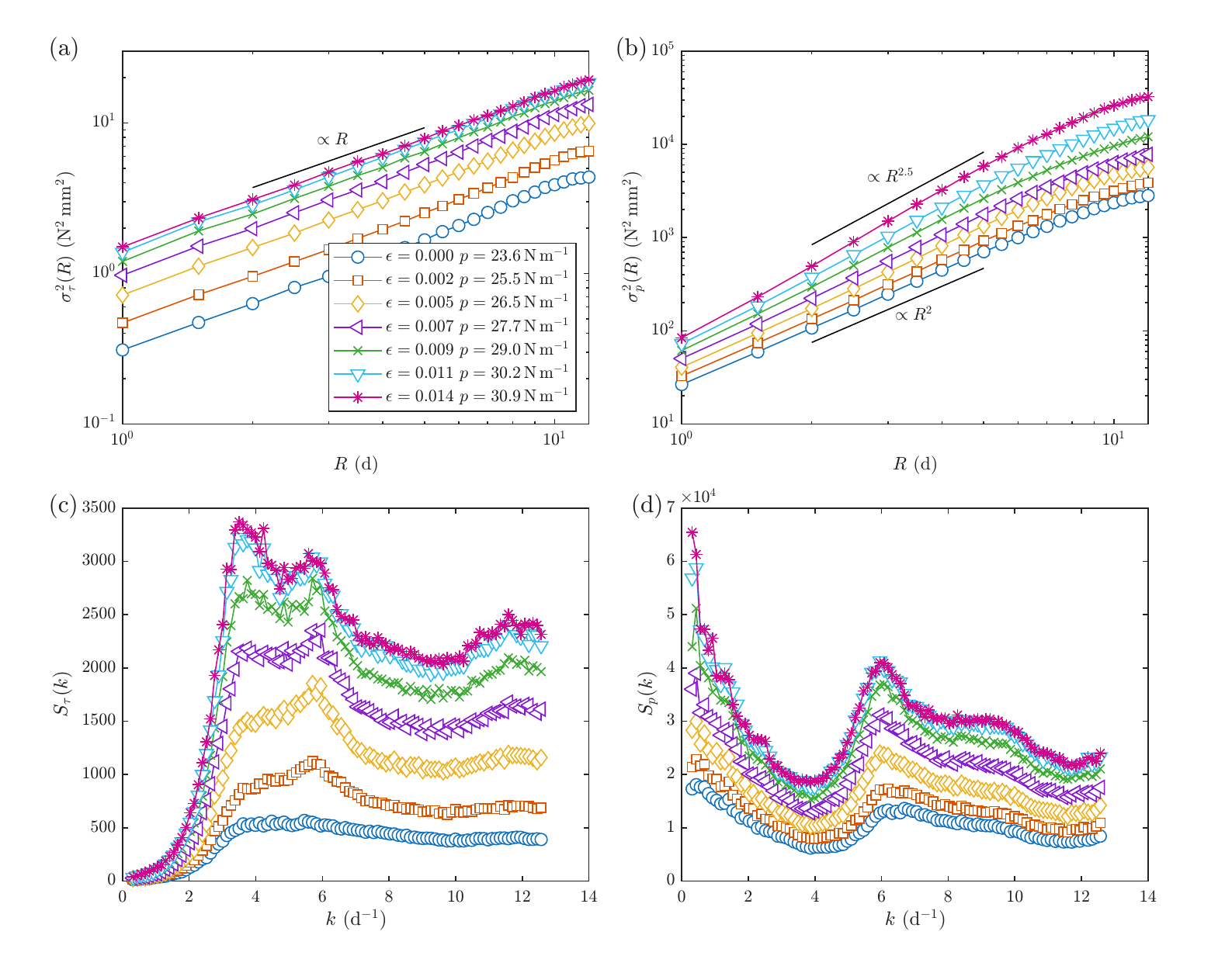}
	\caption{The evolution of the $\sigma^2(R)$ and $S(k)$ curves with strain $\epsilon$ in systems of packing fraction $\phi = 0.859$. }
	\label{fig:f3}
\end{figure}

To better observe the evolution of the data, we present Fig.~\ref{fig:f3}, which shows how the curves change at small strains for a single packing fraction, $\phi=0.859$. It is possible to see the continuous evolution of $\sigma^2_p(R)$ and $S(k)$ during this process. A numerical comparison here might require some form of normalization to counteract the effect of the increasing pressure.

\section{Experiments with Ellipses}
\label{ell}
\subsection{Experimental Setup}

 The experimental apparatus used to study elliptical granules is the one shown in Fig.~\ref{apparat}. Here, we used bi-disperse elliptical particles with two different aspect ratios. The major and minor axes of the small particles are $a=15$ mm, $b = 10$ mm, and $a=20$ mm $b = 10$ mm, respectively. For both particle types, the length ratio of the large to small particles is 1.4:1, and the number ratio is 1:2. This ensured that the area ratio occupied by the large and small particles is approximately 1:1, consistent with the disk setup. The length unit $\ell$ is chosen as the minor axis length of the small ellipse, $\ell = 10\,\mathrm{mm}$.
 
 The isotropic compression protocol is analogous to the one used for disks described in Sec.~\ref{esic}. The main difference is that the jamming point $\phi_J$ is strongly dependent on the aspect ratio $a:b$. All compressions begin from a stress-free, homogeneous packing at an initial packing fraction $\phi_0$ that is near the jamming point. For the $a:b = 1.5:1$ case, $\phi_0 = 0.847$, and for the $a:b=2:1$ case, $\phi_0 = 0.854$.
 
\subsection{Data Analysis}

\begin{figure}
	\centering
	\includegraphics[width= 8.6cm]{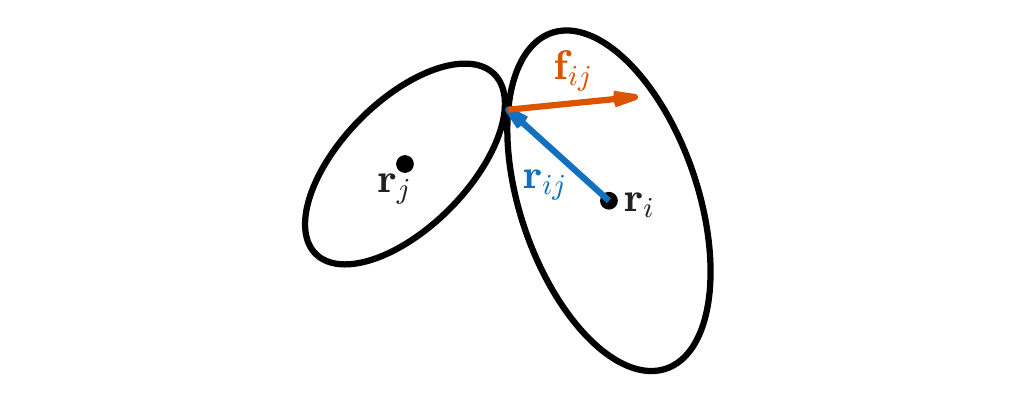}
	\caption{Definitions of $\mathbf{r}_i$, $\mathbf{r}_{ij}$ and $\mathbf{f}_{ij}$ for ellipses.}
	\label{fig:dee}
\end{figure}

For the elliptical particle system, the definitions of $\mathbf{r}_{ij}$ and $\mathbf{f}_{ij}$ are as shown in Fig.~\ref{fig:dee}. Here, $\mathbf{r}_{ij}$ is the vector from the geometric center of ellipse $i$ to the contact point between $i$ and $j$, and $\mathbf{f}_{ij}$ is the interaction force exerted by ellipse $j$ on ellipse $i$. The definitions of the torque field and the pressure field are the same as in Eq.~\ref{eq:tau} and Eq.~\ref{eq:pressure}. A key difference from the circular disk system is that for ellipses, the contact normal is generally not co-linear with $\mathbf{r}_{ij}$. Therefore, the normal force now contributes to the torque, and the frictional force also contributes to the pressure. 

 Once these fundamental fields are calculated, their fluctuations and spectra are analyzed using the same methods as for the disk systems, as described in Eq.~\ref{eq:var} and Eq.~\ref{eq:sk}, respectively.

\subsection{Results}

\begin{figure}[htbp]
	\centering
	\includegraphics[width = 14 cm]{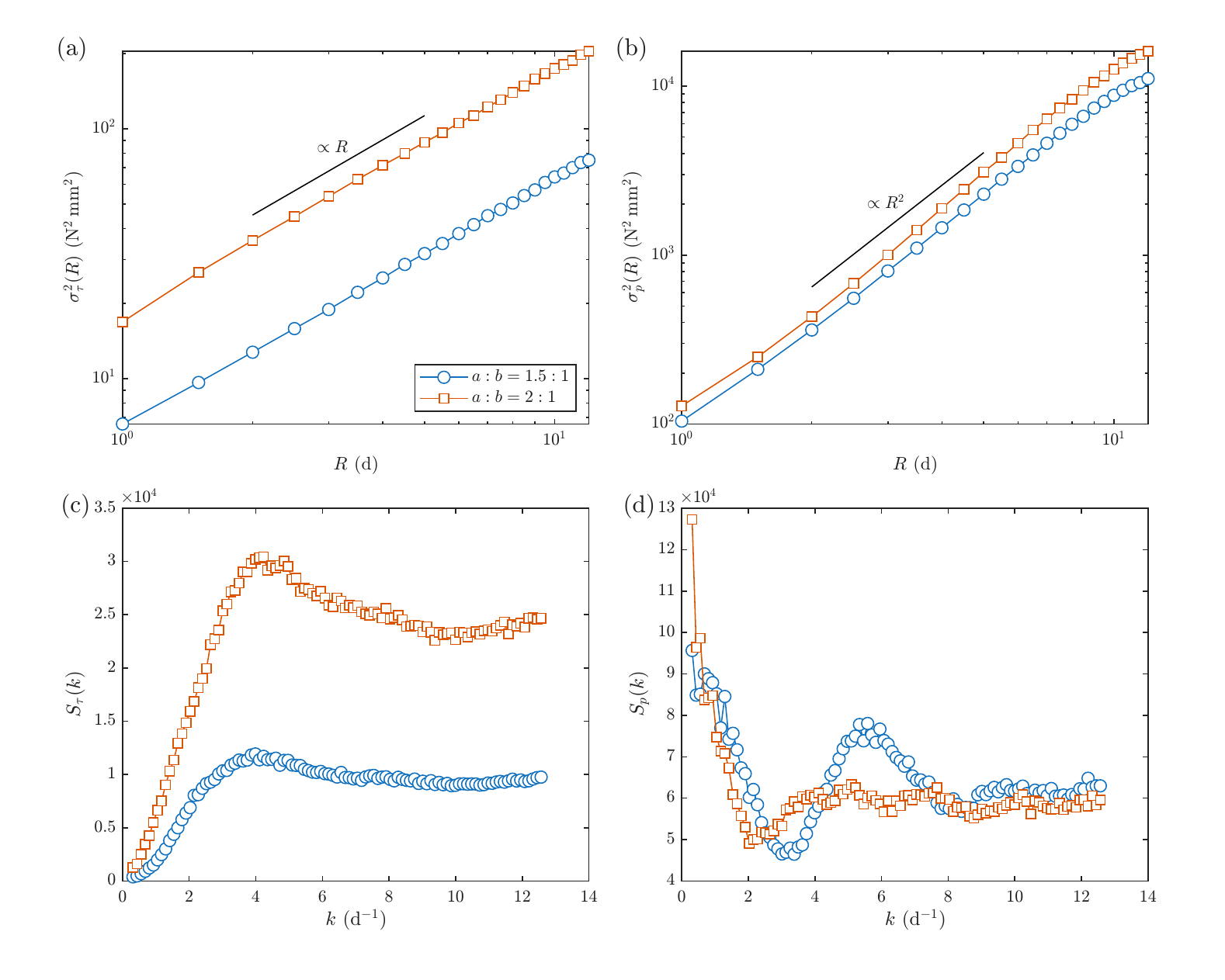}
	\caption{The variance $\sigma^2(R)$ of of torque $\tau$ and pressure $p$ fluctuations and their corresponding $S(k)$ in elliptical systems. }
	\label{fig:f1}
\end{figure}

Fig. \ref{fig:f1} (a) shows the torque fluctuations of the two elliptical systems under isotropic compression. To ensure a fair comparison, we selected systems with similar pressure $p$, both approximately $50\, \mathrm{N\,m^{-1}}$. (The pressure of the previous disk systems was around $30\, \mathrm{N\,m^{-1}}$; since the elliptical and circular particles are made of different materials, the range we can accurately measure also differs.)
The torque fluctuations in both cases exhibited hyperuniformity. The system with an aspect ratio of $a:b = 2:1$ shows larger fluctuations, which is expected because flatter ellipses can generate more torque from normal forces.

Fig. \ref{fig:f1} (b) shows the normal fluctuations of pressure, with very little difference between the two systems.

Fig. \ref{fig:f1} (c) and (d) show the corresponding $S(k)$. For torque $\tau$, the $a:b=2:1$ system appears to be more linear at small $k$. For $p$, there are some differences at large $k$, which may be due to the effect of the particle shape on the contact structure. 

\section{Summary and Discussion}
\label{summary}
\begin{figure}
	\centering
	\includegraphics[width= 8.6cm]{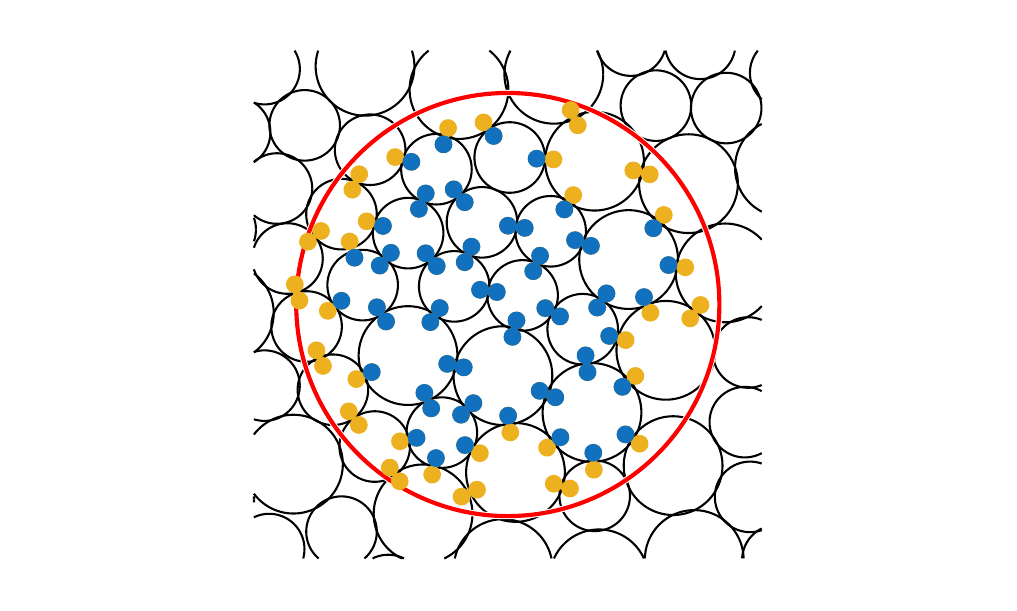}
	\caption{Illustrating the consequences of the constraint of the vanishing of the torque for each particle. See text for explanation}
	\label{constraint}
\end{figure}

The upshot of the experimental results presented above is that the hyperuniformity of the torque fluctuations is very robust, and in all the cases examined above the torque variance increased linearly in $R$. The scaling of the torque variance remains unchanged even when the pressure variance deviates from normality due to the loss of system's isotropy. This robustness requires an explanation. Indeed, the reason for this is quite intuitive, and results from the constraint on the resultant torque which must vanish on each particle. The consequences of this constraint can be seen in Fig.~\ref{constraint}. Integrating the torque inside any randomly chosen circle (outer red line), one gets no contribution from the inner contacts (highlighted as pairs of blue points), since they are totally within the circle, they contribute exactly zero. The only contribution comes from contacts that are adjacent to the outer boundary, where some contacts are within (highlighted  as pairs of yellow points) and some are outside the circle (not highlighted ). Since the circle cuts the configuration randomly, the values of torque computed along the boundary are random, and will exhibit a normal variance increase proportional to the length of the boundary, in agreement with the observation presented in this paper. 

Although our experiments are conducted in 2-dimensions, we can use the intuitive argument to generalize these results and predict that in d-dimensions the torque variance will increase like $R^{d-1}$, being hyperuniform as well. While at this time it seems quite difficult to confirm this prediction in experiments conducted in 3-dimensions, it appears worthwhile to test the prediction in numerical simulations of frictional matter in 3-dimensions.

\acknowledgments 

This research was pursued at the Hangzhou International Innovation Institute, Beihang University, Hangzhou, China, without whose hospitality it could not have been conducted.  JS and JZ thank Yinqiao Wang for helpful discussions. JS and JZ acknowledge the support of the NSFC (Nos. 11974238 and 12274291) and also the support of the Innovation Program of Shanghai Municipal Education Commission under No. 2021-01-07-00-02-E00138. JS and JZ also acknowledge the support from the Student Innovation Center of Shanghai Jiao Tong University.

\bibliography{All}

\end{document}